# Near-Field Coupling Coil System: A Novel Radiofrequency Coil Solution for MRI

Zhiguang Mo[1,2], Shao Che[3], Enhua Xiao[1,2], Qiaoyan Chen[1,2], Feng Du[1,2], Nan Li[1,2], Sen Jia[1,2], Changjun Tie[1,2], Bing Wu[3], Xiaoliang Zhang[4], Hairong Zheng[1,2], Ye Li[1,2]*

[1]*Paul C. Lauterbur Research Center for Biomedical Imaging, Shenzhen Institutes of Advanced Technology, Chinese Academy of Sciences, Shenzhen 518055, China.
[2]*Shenzhen Key Laboratory for MRI, Shenzhen 518055, China.
[3]Shanghai United Imaging Healthcare Co., Ltd., Shanghai, China.
[4]Department of Biomedical Engineering, State University of New York at Buffalo, NY, United States., Buffalo, NY, United Stats.

*Corresponding author. E-mail: liye1@siat.ac.cn

## Abstract

The performance of radiofrequency (RF) coils has a significant impact on the quality and speed of magnetic resonance imaging (MRI). Consequently, rigid coils with attached cables are commonly employed to achieve optimal SNR performance and parallel imaging capability. However, since the adoption of MRI in clinical imaging, both patients and doctors have long suffered from the poor examination experience and physical strain caused by the bulky housings and cumbersome cables of traditional coils. This paper presents a new architectural concept, the Near-Field Coupling (NFC) coil system, which integrates a pickup coil array within the magnet with an NFC coil worn by the patient. In contrast to conventional coils, the NFC coil system obviates the necessity for bed-mounted connectors. It provides a lightweight, cost-effective solution that enhances patient comfort and supports disposable, custom designs for the NFC coils. The paper also derives the SNR expression for the NFC coil system, proposes two key design principles, and demonstrates the system's potential in SNR and parallel imaging through an implementation case.

# Introduction

Magnetic resonance imaging (MRI) is a non-ionizing imaging modality that provides high-resolution anatomical images of the human body through non-invasive scanning. It has become an indispensable tool for clinical diagnosis and basic biomedical research. Over the past few decades, the design of magnetic resonance radiofrequency (RF) coils has primarily focused on enhancing SNR performance [1], as higher SNR allows for obtaining high-contrast, high-resolution images [2,3,4,5,6]. To achieve this, ultra-flexible coil arrays [7,8] and high-density RF coil [9] arrays have been developed. The former, while offering improved signal-to-noise ratio (SNR) due to its closer proximity to the imaging target area, faces challenges related to its flexibility and examination convenience due to the presence of components such as preamplifiers and cables. The latter, on the other hand, involves an increasing number of RF cables and connectors due to the higher number of channels [10]. As these high-density coil arrays approach their theoretical limits at a given field strength [11], the challenges related to the bulk and weight of the coils and cables become increasingly evident. These cables are designed to be long and thick to accommodate a variety of positioning needs and ensure reliability and low losses. To mitigate surface-induced currents on the cables during system transmission, multiple bulky RF traps are added. When changing the scan area, the procedure involves disconnecting all connectors, replacing the coils—each weighing several kilograms, such as those used for the head or knee—and then reconnecting the connectors. Additionally, the cables present on the patient bed can pose a risk of burns to patients due to faults or poor design, which may also contribute to their increased anxiety. Examples of traditional head and knee coils are shown in Figure 1A.

Researchers have made multiple attempts to eliminate cables. Wiggins and colleagues tried to replace all dedicated coils with a 128-channel bore-liner coil integrated within the magnet. Unfortunately, they encountered unexpected practical challenges, and this "universal" remote body array failed to become a practical option for routine imaging [12]. Another approach involved integrating additional "on-coil" communication antennas and associated electronic devices into the receive coil to replace the cables [13,14,15]. However, these methods increased the complexity of the system, introducing new challenges related to the wireless power supply for integrated electronic devices and the synchronization of clock signals [16].

As early as 2001, research demonstrated that resonant structures could effectively guide RF flux from an object to a remote receiver coil [17]. Since then, these structures have been extensively employed over regions of interest (ROIs) to enhance the SNR of the receive coil [18,19,20,21,22]. Recently, there has been a growing trend to incorporate resonant structures into the development of wireless coil solutions [23,24]. Among these, a birdcage-shaped wireless resonant structure, combined with body coils, has been used to achieve signal reception [25]. While this structure performs effectively for imaging at low field strengths and small objects like the wrist [25], its SNR at 3T field strength or in large field-of-view conditions is typically lower compared to traditional coils [23,26,27].

In comparison to birdcage-shaped structures, decoupled multi-element coil arrays generally offer superior SNR performance, which is the primary reason why traditional receiving coils are often designed as decoupled arrays. Recently, decoupled wireless coil arrays have shown significant potential for enhancing SNR performance [28,29,30]. However, the aforementioned solutions all use a body coil or partial spine coil elements as receive coils, leading to limited parallel imaging capability [25,26,27,28], which is crucial for achieving high-efficiency magnetic resonance scanning in clinical applications.

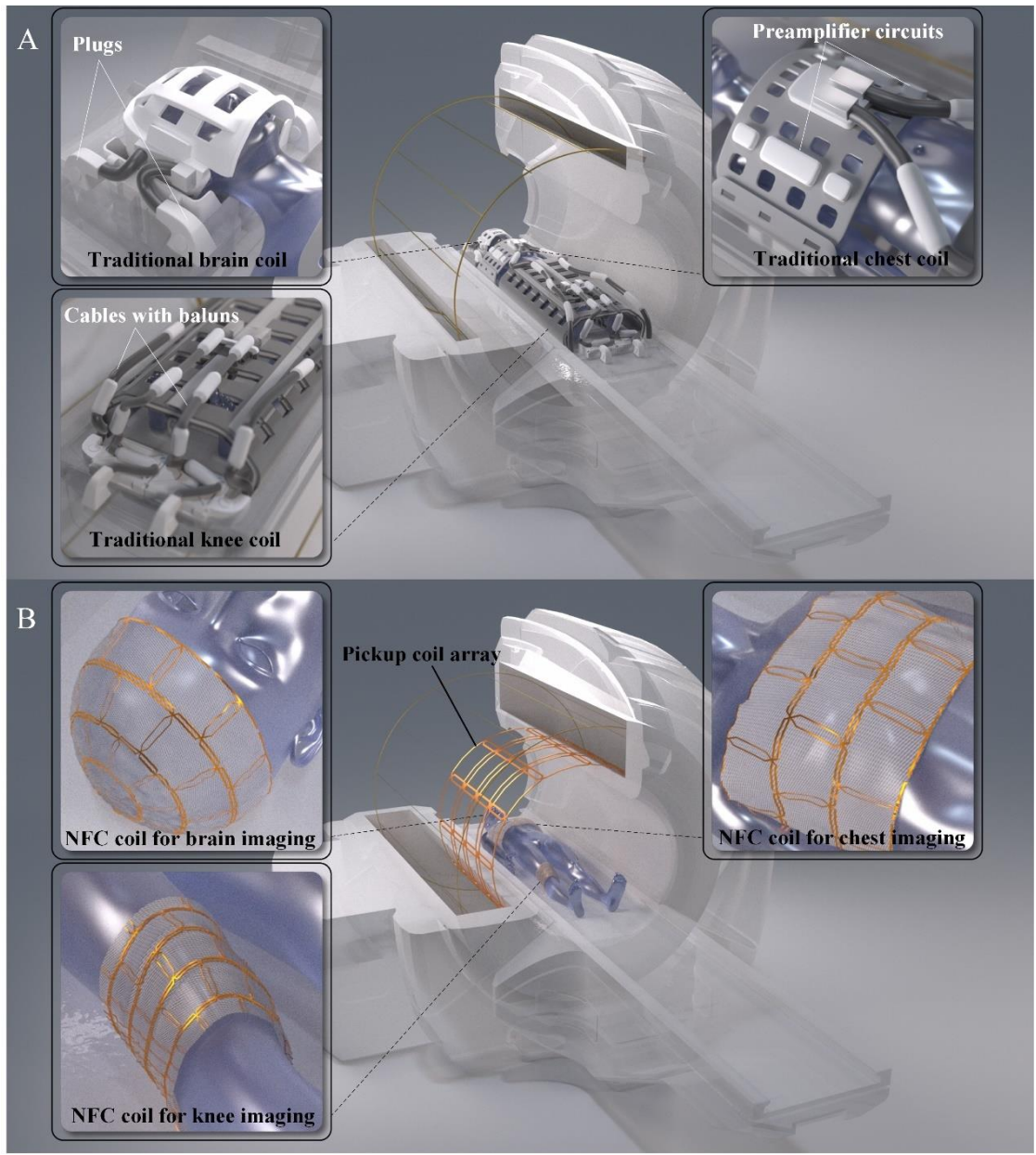

**Figure 1. The schematic diagrams of the MRI RF coil.** (A) Schematic of a traditional RF coil system. (B) Conceptual diagram of the NFC coil system. Some electronic components are omitted in the figures.

In this paper, we present the concept of the NFC coil system, which consists of two main components: an NFC coil comprising multiple decoupled resonant elements and a bore-liner pickup coil array. The conceptual diagram of the NFC coil system is shown in Figure 1B. The use of the bore-liner pickup coil provides superior SNR compared to using a body coil, thereby overcoming the limitations of body coils in parallel imaging due to their limited channel count and the spatial constraints of spine coils, which are confined to a single plane. This paradigm-shift NFC coil system technique enhances both the SNR and parallel imaging capabilities. The Performances of the NFC coil system compared to traditional coil arrays, flexible coil arrays, and the integrated RF/wireless coil (iRFW) [16, 31] are presented in Table 1. To validate our concept, we derived the SNR expression for the NFC coil system and confirmed its accuracy through simulations and experiments. Subsequently, we developed an NFC coil system and compared its SNR and parallel imaging capabilities with those of its pickup coil and a commercial knee coil through in vivo studies. The results suggest that the proposed NFC coil system has the potential to meet the clinical needs for SNR and parallel imaging capabilities, making it a promising new solution for wireless coils.

Table 1: Performance Comparisons of Different RF Coil Types

| Types | Channel Count | Body Conformity | Preamp-free | Cable-free | No need additional modules | Per-channel electronic component weight (g) | Per-channel electronic component costs (USD) |
|---|---|---|---|---|---|---|---|
| Traditional Rigid coil array | Midian | Low | No | No | - | ≈ 100 | ≈ 230 |
| Flexible coil array | Low | Midian | No | No | - | ≈ 100 | ≈ 230 |
| iRFW | Low | Low | No | Yes | No | ≈ 110 | ≈ 250 |
| NFC coil (This work) | High | High | Yes | Yes | Yes | ≈ 5 | ≈ 5 |

# Results

**Design principle 1: maximizing the detuning performance for NFC coil elements**

The NFC coil elements must remain detuned during the transmission phase to prevent the transmission energy from being directly captured and amplified by the resonant elements, which could otherwise interfere with the transmission field. Currently, there are two primary circuit structures developed to achieve the requirements: the "bypass capacitor circuit" [22, 25] and the "parallel resonance circuit."[29, 30, 32, 33] To compare the detuning effects of these two circuit structures and their impact on imaging, we fabricated NFC coil elements using each detuning circuit structure, as illustrated in Fig. 2A and 2B.

During the reception phase, the RF signals from the human body are insufficient to forward-bias the bidirectional diodes. Consequently, the equivalent circuits of the two coil elements are identical. By adjusting the tuning capacitors C2 and C3, the resonance frequency of the coil elements can be tuned over a broad range, depending on the values of the tuning capacitors. Fig. 2C shows that by varying these capacitors, which range from 4.7 pF to 20 pF, the frequency can be adjusted in the range of 102 MHz to 211 MHz, sufficiently accommodating any shifts in the resonant frequency.

During the transmission phase, the electromagnetic waves emitted by the transmit coil generate an electromotive force (EMF) across the ends of the detuning circuit, causing the bidirectional diodes to conduct. At this juncture, the capacitance in the "bypass capacitor circuit" is bypassed, resulting in a shift in the resonant frequency of the coil element away from the operating frequency. As a result, the coil element exhibits minimal resonance at the operating frequency, but it hardly imposes any impedance on the current signal on the coil. Conversely, in the "parallel resonant circuit," the L1 and C2 components form a parallel resonance, thereby impeding the flow of current through the coil. As shown in Figure 2D, after disconnecting the capacitors C1 and C3, and shorting the grounds of the two ports of the vector network analyzer, the center conductors of the coaxial cables are connected to the respective ends of the two detuned circuits (dual-port method).

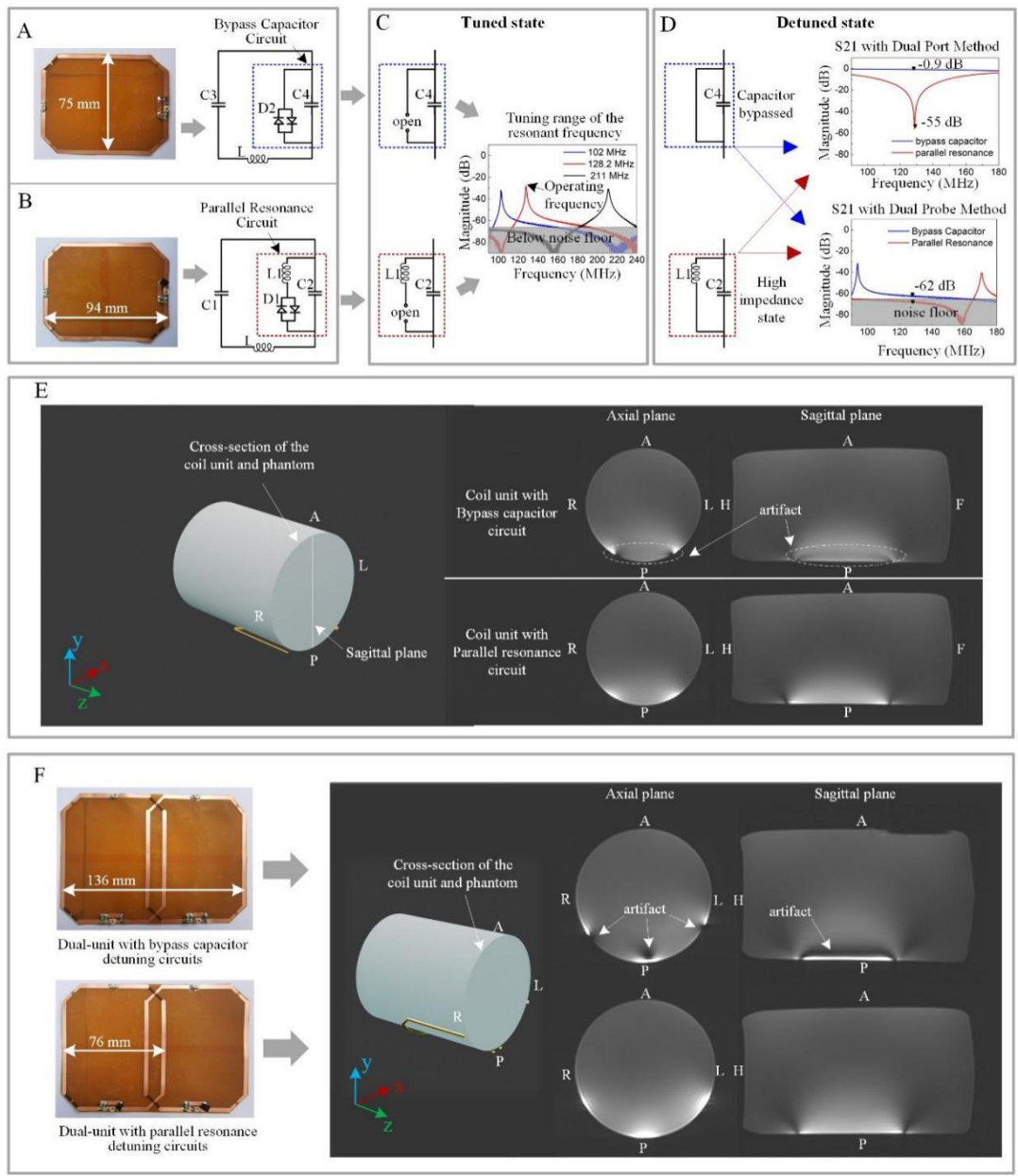

**Figure 2. NFC coil elements employing parallel resonance method and bypass capacitance method, respectively.** (A) Coil element with parallel resonance detuning circuit. (B) Coil element with bypass capacitance detuning circuit. (C) Equivalent circuit and frequency tuning range of coil element in resonant state. (D) Equivalent circuit diagram of the coil element in detuned states using two detuning methods, along with the transmission coefficient of the two coil elements measured in detuned state using a dual-port, and their magnetic response measured with a dual-loop. (E) Axial plane and sagittal plane MRI images of a phantom obtained using the two NFC coil elements. (F) The Dual-element NFC coils using the two detuning circuits separately, along with the MR images obtained using each of them.

Under conditions where the diodes are fully conducting, the transmission coefficients of the two detuned circuits are observed to be -0.9 dB and -55 dB, respectively. A dual-probe consisting of two decoupled wideband loops, with an isolation of -65 dB, was connected to a vector network analyzer to evaluate the response of the two NFC coil elements to an external electromagnetic field (dual-probe method). The transmission coefficient of the coil element using the "bypass capacitor circuit" as an intermediary was measured to be -62 dB, while that of the "parallel resonant circuit" approached the noise floor of the dual-probe. These results suggest that the parallel resonant circuit provides significantly enhanced signal attenuation compared to the bypass capacitor circuit.

To further assess the impact of detuning performance on MRI, we employed two types of NFC coil elements along with the pickup coil shown in Fig. 1A and 1B in phantom imaging experiments. The cross-sectional and sagittal MRI images exhibit artifacts when the NFC coil element with the 'bypass capacitor circuit' is used, while the NFC coil element with the more effective detuning provided by the 'parallel resonant circuit' shows no significant artifacts (Fig. 2E). When the number of elements is increased to two, the artifacts in imaging caused by insufficient detuning in the NFC coil with the bypass capacitor scheme become more pronounced (Fig. 2F).

**Design principle 2: minimizing coupling between NFC coil elements**

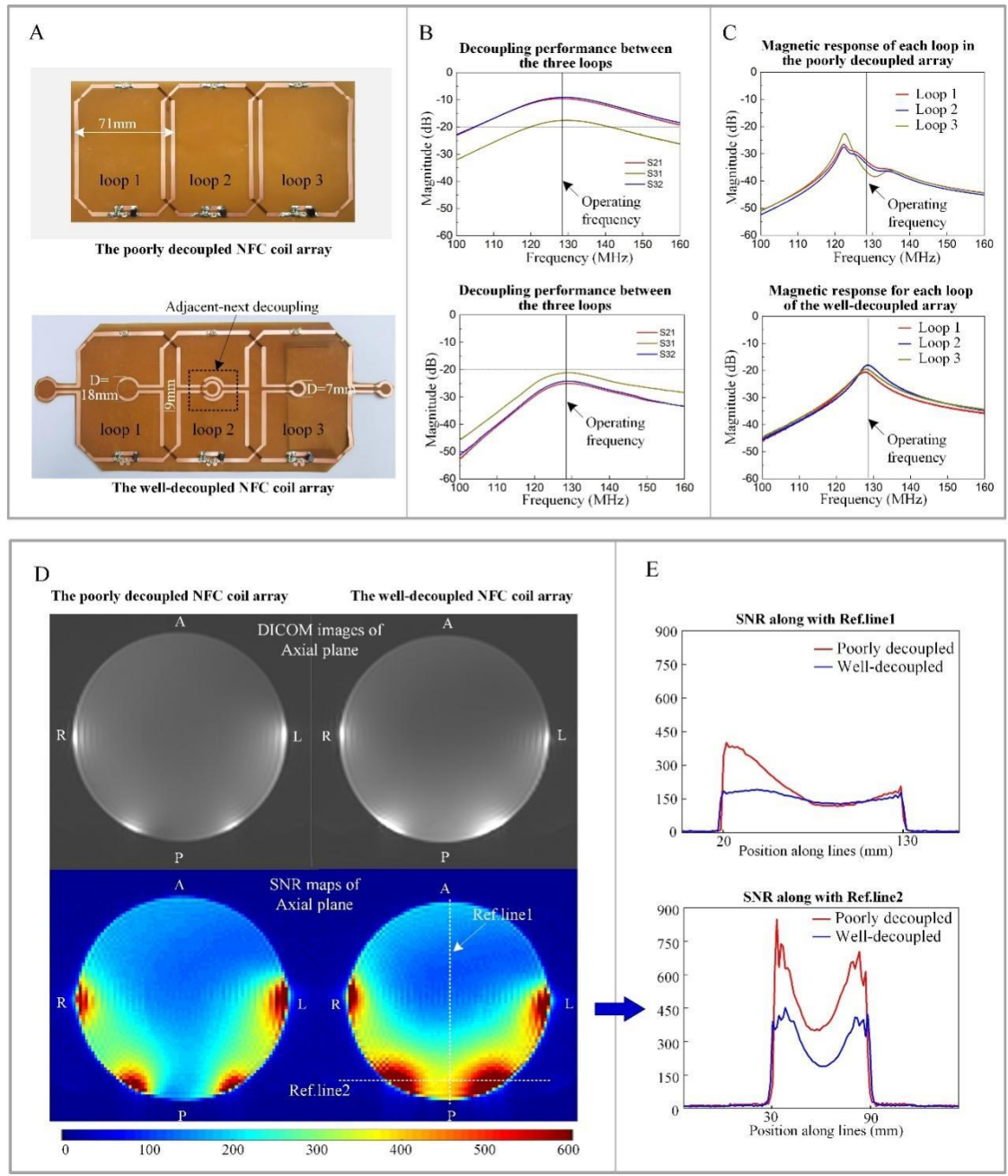

**Figure 3. The impact of inter-element decoupling performance on the imaging SNR of NFC coils.** (A) Poorly decoupled and well-decoupled NFC coils. (B) In the poorly decoupled NFC coil, the coupling coefficient between adjacent elements exceeds -10 dB, and the next-nearest elements exceed -18 dB. Each element's resonant frequency, measured by dual-probe, deviates from the operating frequency. (C) In the well-decoupled array, the coupling coefficients between adjacent and next-nearest elements are all below -20 dB, ensuring that the resonant waveform of each element remains unaffected by the other elements. (D) MR images and SNR maps obtained using poorly decoupled and well-decoupled NFC coil, respectively. (E) Comparisons of SNR obtained with the two NFC coils along reference lines 1 (Ref.line1) and reference line 2 (Ref.line2).

In the traditional wired RF coil design, when the number of channels increases from one to two, the signal voltage doubles while the equivalent noise voltage increases by a factor of $\sqrt{2}$, theoretically boosting the SNR by a factor of $\sqrt{2}$ [34]. A common method to improve SNR is to increase the density of RF coil coverage in the region of interest. However, this approach requires sufficiently low coupling between the channels to ensure that the signals and noise from the two channels remain uncorrelated. Similarly, in the proposed NFC coil system concept, it is crucial to minimize the coupling between the elements of the NFC coils. In this context, it is important to note that "decoupling" refers to the coupling between NFC coil elements, whereas "coupling" in the near-field coupling coil system pertains to the interaction between the NFC coil and the pickup coil array. This has two main advantages: first, it can potentially improve SNR performance; second, when the NFC coil elements are expanded into a multi-element array, low coupling ensures that the resonant state of each element remains unaffected by the others, allowing for independent tuning of the resonant frequencies.

To evaluate the negative impact of inadequate decoupling between channels on the SNR of a NFC coil, we fabricated two three-element NFC coils in both poorly decoupled and well-decoupled configurations, as shown in Fig. 3A, with each element tuned independently to the system's operating frequency. In the poorly decoupled NFC coil, the coupling coefficient between adjacent elements exceeds -10 dB, while the coefficient between the next-nearest elements exceeds -18 dB. This level of coupling results in the resonant frequency of each element in the poorly decoupled array to deviate from the operating frequency of the system (Fig. 3B). Conversely, in the well decoupled array, the coupling coefficients between both adjacent and next-nearest elements are below -20 dB, ensuring that the resonant waveform of each element remains undisturbed by the others (Fig. 3C). Magnetic resonance imaging with these two NFC coils demonstrated that the well-decoupled array provides a notable SNR advantage, as shown in both DICOM images and SNR maps (Fig. 3D). The SNR curves along the two reference lines highlight that this advantage is particularly evident in areas near the NFC coil (Fig. 3E).

**Phantom experiments: SNR and B1 field distribution of the NFC coil system**

As a complex system comprising a multitude of pickup and NFC coil elements, the NFC coil system may give rise to curiosity and concern regarding the underlying signal transmission mechanisms. To address this issue, we derived an SNR expression for the NFC coil system when combining $n$ NFC coil elements with $m$ pickup coils. The formula shows that the signal distribution in the NFC coil system is a weighted combination of the B1 fields from all NFC elements, with the weights determined by the mutual inductance coefficients between each NFC coil and all pickup coils.

To validate this formula, a 6-element NFC coil was constructed based on the aforementioned theoretical principles. A row of four spine channels was used as pickup coils to conduct imaging experiments on an 11 cm diameter phantom (Fig. 4A). A schematic diagram of the cross-sections of the NFC coil elements and spine coil is shown in Fig. 4B. The mutual inductance coefficients between each pickup coil element and each NFC coil element were calculated using 3D electromagnetic simulation software (CST Microwave Studio Suite), and are displayed in Fig. 4C. The final column highlights the NFC coil elements with higher mutual inductance coefficients, which, according to the formula, will primarily influence the NFC coil system sensitivity map with their B1 field, whether using a single pickup coil element or a combination of all pickup coil elements. This is corroborated by the results shown in Fig. 4D, where the dominant NFC elements identified in the MRI images, SNR maps, and simulated B1 fields from the phantom study align with those listed in the table in Fig. 4C, regardless of whether a single pickup coil or a combination of all pickup coils is used.

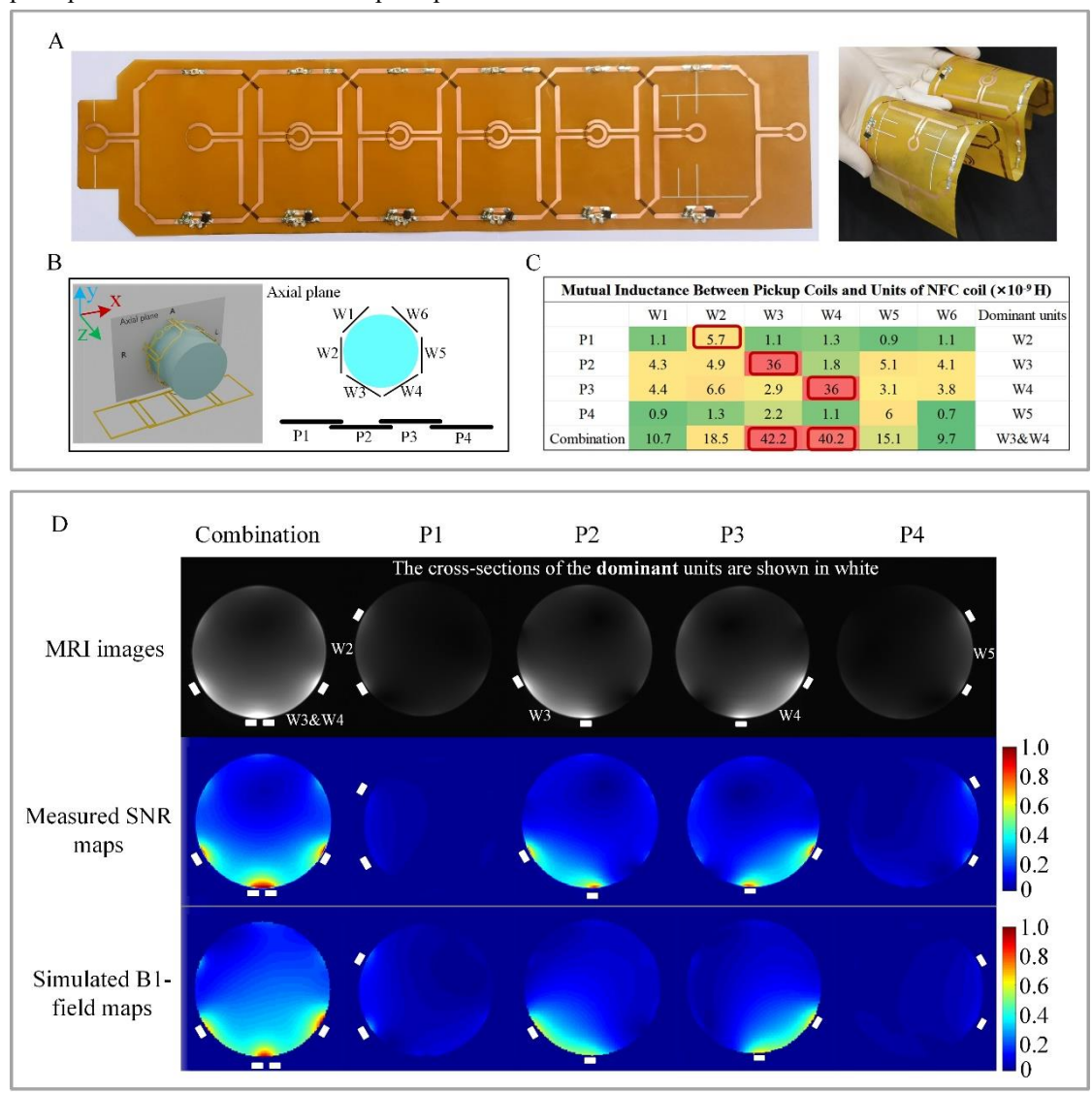

**Figure 4. Phantom-based MRI validations for the sensitivity combination of the NFC coil system.** (A) The photos of the 6-element NFC coil (B) Axial plane images of a phantom using a 4-channel spine coil array and a 6-element NFC coil. P1-P4 represent images obtained using a single pickup coil element, while "combination" represents the image obtained using all four pickup coils. (C) Mutual inductance coefficients using 3D electromagnetic simulation software between each channel of the four pickup coils and the six NFC elements. (D) MRI images, measured SNR maps, and simulated B1- maps show the combined sensitivity and per-channel sensitivity distribution of the NFC coil system. SNR maps and B1-field have both been normalized to their maximum values. The dominant NFC elements shown in the images and maps are consistent with the dominant NFC elements listed in the mutual inductance coefficient table.

**In vivo experiments 1: the validation of NFC coil system SNR performance**

The SNR expression for the NFC coil system indicates that when the weight coefficients of the B1 fields from each NFC element are similar, its SNR expression is essentially consistent with that of a wired coil array of the same size. This suggests that a well-designed NFC coil system has the potential to achieve a better signal-to-noise ratio in comparison to the rigid coils currently used in clinical practice, as rigid coils are often designed to accommodate extreme sizes and may not fit the dimensions of most individuals optimally.

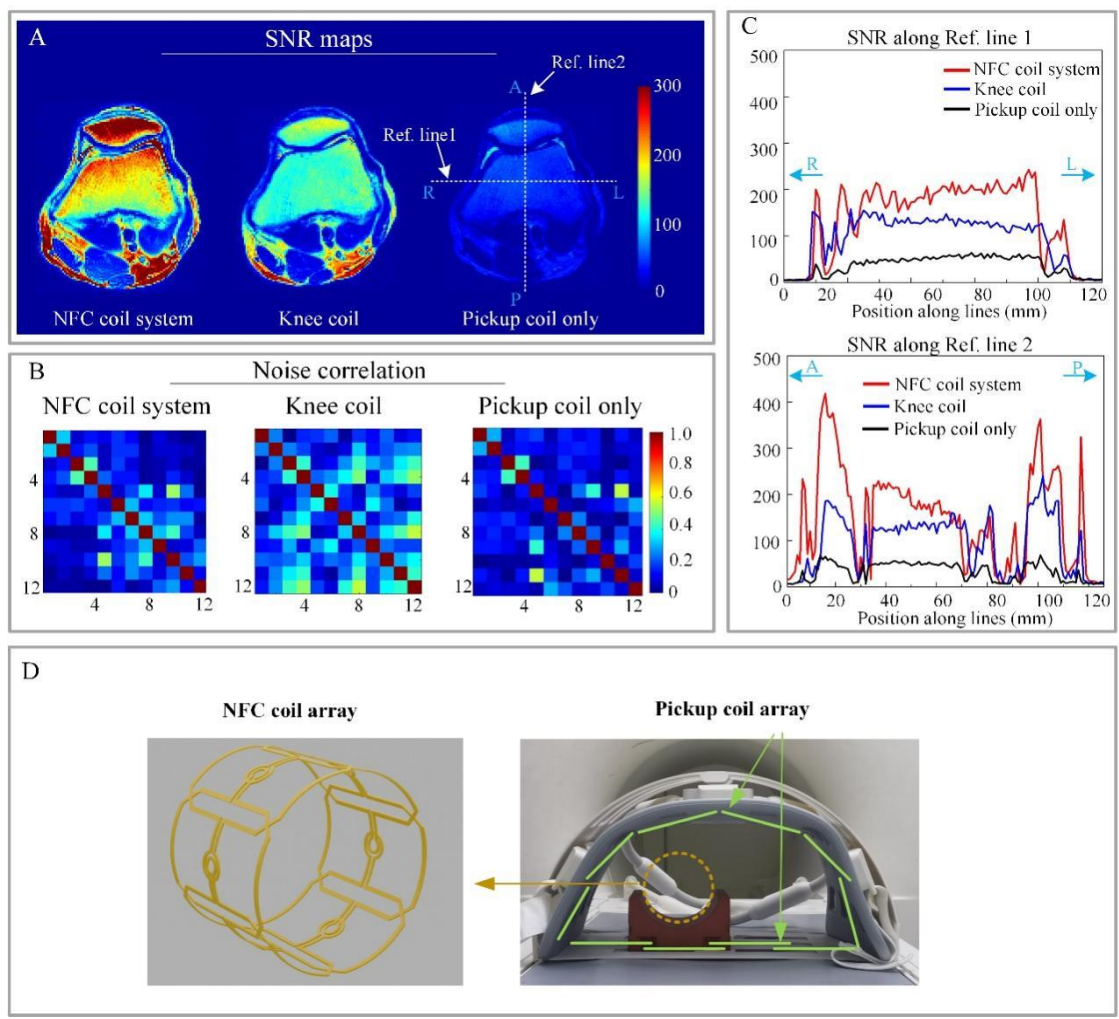

**Figure 5. Comparison of the SNR performance of NFC coil system, its pickup coil, and a commercial knee coil.** (A) The SNR maps obtained using the NFC coil system, its pickup coils and the knee coil. (B) The noise matrices used for calculating the signal-to-noise ratio (SNR). (C) The SNR comparisons along reference line 1 (Ref. line1) and reference line 2 (Ref. line2). The SNR of the images obtained along the reference line using NFC coil system improved by approximately 360% compared to the pickup coil and by about 60% compared to the knee coil. (D) The NFC coil and pickup coil used for acquiring the SNR map.

To validate the perspective, a 6-element NFC coil was customized and tailored to the dimensions of the volunteer's knee, in accordance with the two aforementioned design principles. Furthermore, magnetic resonance imaging of a volunteer's knee was conducted using the NFC coil system, its pickup coil, and a commercial rigid knee coil. The SNR maps in the axial plane were calculated using the "cov-rSOS" method [35], and the SNR curves were plotted along two reference lines (Fig. 5A). The average SNR values along these lines in the knee region indicate that the NFC coil system improves the SNR by 360% in comparison to its pickup coil and by 60% in comparison to the knee coil (Fig. 5C). The noise covariance matrix used to calculate the SNR indicates that the NFC coil system does not significantly worsen in noise performance in comparison to the standalone pickup coil (Fig. 5B). The SNR performance allows the NFC coil system to achieve imaging with a resolution of up to 0.33 mm × 0.33 mm × 2 mm. The MRI images obtained using the NFC coil system and the knee coil are shown with the same window width and level in Fig. 5D.

**In vivo experiments 2: the parallel imaging capability of NFC coil system**

Parallel imaging technology improves scanning speed through the utilization of multiple receive coil elements, which enables the capture of localized spatial information, thereby reducing the number of required acquisitions [36, 37].

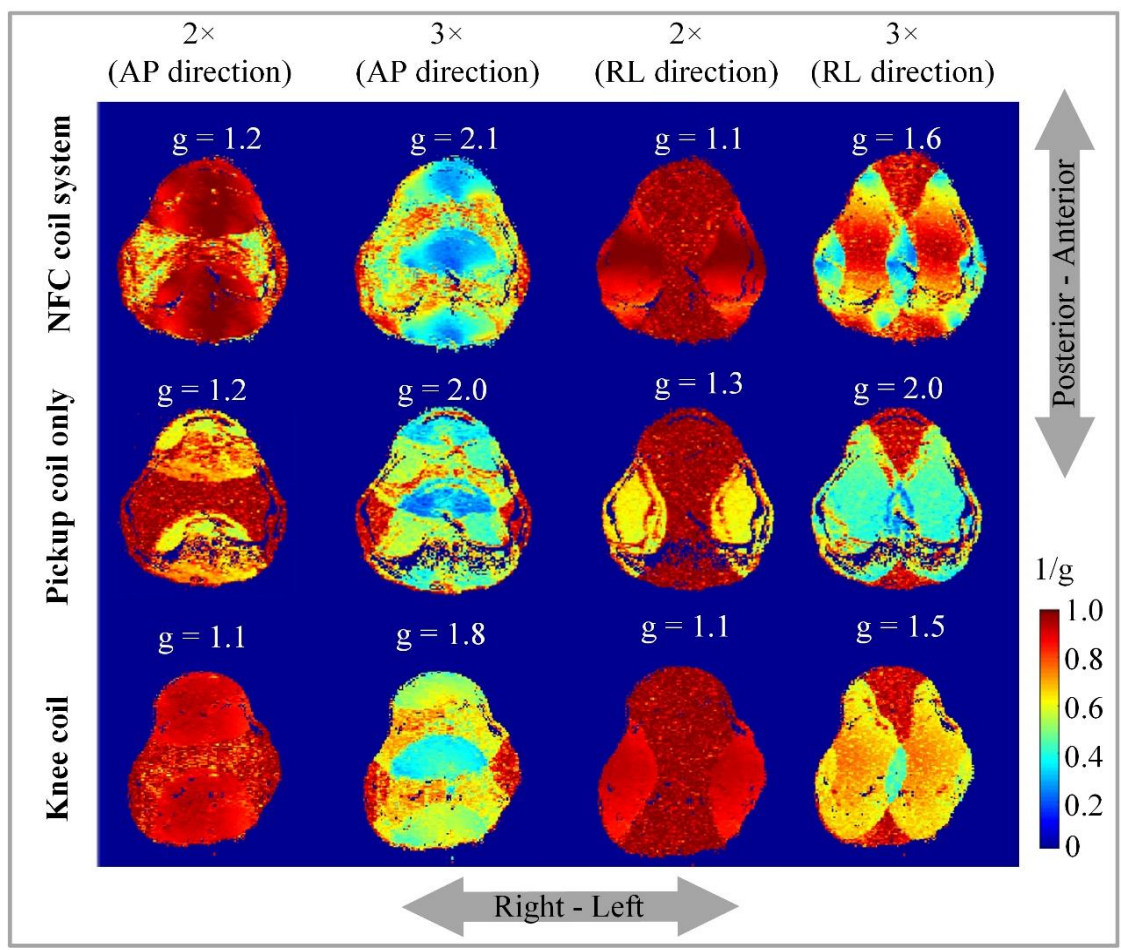

**Figure 6. Comparison of the parallel imaging capabilities of NFC coil system, its pickup coil, and the knee coil.** The figure shows the mean g-factor (highlighted in white, where a smaller g-factor corresponds to better acceleration performance) and the 1/g maps under the same acceleration settings.

This necessitates the availability of a sufficient number of receive coil channels. The use of body coils, which typically have only two channels, and spine coils, which, although having more channels, are restricted to a single plane, presents a challenge in achieving robustly accelerated imaging. This is also due to the fact that spine coils are unable to capture information from other directions of the imaging subject. The NFC coil system design presented in this paper employs a total of twelve coil elements as pickup coils, comprising three spine coil elements and nine surface coil elements.

To validate the parallel imaging capability of the NFC coil system, axial imaging of a volunteer's knee was performed using the NFC coil system, its pickup coils, and a knee coil (Fig. 6). Parallel imaging was conducted with acceleration factors of 2 and 3 in the AP and RL directions, respectively. The g-factor, as calculated using the SENSE [37] algorithms, is a commonly used metric for assessing acceleration performance, with a smaller g-factor denoting better performance. The colored images, which represent the 1/g maps, provide an intuitive visualization of the data.

The results indicate that the parallel imaging capability of the NFC coil system in the RL direction is significantly superior to that of the pickup coil, while it is comparable to that of the knee coil. This suggests that with the same number of pickup coil channels as traditional coils, the parallel imaging capability of the NFC coil system has the potential to reach the same level. On the other hand, in the AP direction, the g-factor of the NFC coil system is slightly inferior to that of the knee coil, which may be attributed to the lower density of coil elements in the AP direction for the pickup coil.

## Discussion

In this study, we have introduced the concept of the Near-Field Coupling (NFC) coil system to provide a reference paradigm for the next generation of wireless RF coils. Unlike traditional coils connected to the patient bed via cables, the NFC coil system facilitates signal acquisition through a combination of a patient-worn NFC coil and a bore-liner pickup coil array integrated into the magnet. The NFC coil system eliminates the need for frequent replacement (Fig. 1B) without compromising SNR or parallel imaging performance. For different anatomical applications, only the NFC coil needs to be designed accordingly. Owing to its simple structure—free of the bulky cable assemblies and additional RF components (e.g., baluns, cable traps) required by traditional coils—the NFC coil can more effectively conform to the region of interest, thereby maximizing SNR performance. Meanwhile, the high-density pickup coils allow the NFC coil system to achieve parallel imaging capabilities that are on par with those of conventional coils. The simple structure of the NFC coils also offer a lightweight design and come at a lower cost, which enhances patient comfort and enables customization of the NFC coils, making them disposable. Customization ensures that the NFC coils can adapt to each individual examination site, maximizing SNR performance, while disposability allows the examination to be performed in a more hygienic environment.

To validate the feasibility of the NFC coil system concept and address concerns about potential interference between multiple pickup coil elements and multiple NFC coil elements, which could affect performance, as well as to provide design guidance for the NFC coil, we derived the SNR formula and implemented a 6-element knee imaging case based on two design principles. In vivo studies demonstrated that the NFC coil system achieved a 360% improvement in SNR compared to its pickup coils and a 60% improvement compared to a traditional knee coil. Additionally, parallel imaging experiments showed that the NFC coil system offers parallel imaging capabilities superior to those of its pickup coils and comparable to those of the commercial knee coil.

In the design of coils for clinical applications, SNR performance and parallel imaging capability are the two most critical factors. The former contributes to high-contrast, high-resolution MRI images, while the latter allows for reduced scan times, which is significant for reducing SAR deposition, enhancing patient comfort, and alleviating the pressure on MRI resources.

In terms of SNR, one of the primary contributions of this work is the derivation and validation of the SNR expression for an $m \times n$ element NFC coil system. This expression reveals two key insights: (1) The signal distribution in the NFC coil system is determined by the weighted combination of the B1 fields from each NFC coil element, with the weighting coefficients based on the mutual inductance between each NFC element and all pickup coil elements. (2) When the B1 field coefficients of the NFC elements are equal, the SNR expression for the NFC coil system closely resembles that of a wired coil array of the same dimensions. The first point indicates that when designing an NFC coil system, it is beneficial to draw upon the design principles of wired coil elements and adjust the diameter of the NFC coil to ensure adequate B1 field penetration across the entire knee. This approach is exemplified by the

NFC coil elements used in this study, which were designed with diameters that are appropriately matched to the size of the knee. The second point suggests that the NFC coil system may offer an SNR advantage over rigid coils that do not conform well to the shape of the subject. This is corroborated by the results of our in vivo study, which showed a 60% increase in SNR with the NFC coil system in comparison to the rigid knee coil. It is worth noting that this formula is broadly applicable to a range of decoupling techniques employed between NFC coil elements. It applies not only to the overlapping decoupling approach described in this paper but also to high-impedance decoupling as outlined in reference [28] and analogous techniques, provided that the distance between the pickup coils and the phantom exceeds the radius of the pickup coils and that conductor losses on the pickup coil are negligible.

Additionally, NFC coil system suggests the utilization of a decoupled multi-element resonator array for the NFC coil, as descripted in this paper and in Reference [30]. In contrast, coupled resonators, such as those employing birdcage structures, are better suited for low-field applications and for imaging small body parts, such as the wrist [25]. At 3T field strength or with a large FOV, their signal-to-noise ratio (SNR) is observed to be comparatively lower than that of conventional coils [23, 26, 27]. According to the SNR formula, this phenomenon may be attributed to a factor analogous to that observed in wired coils. A decoupled multi-channel element array typically exhibits a lower level of sample noise in comparison to structures such as the birdcage resonator, thereby providing an advantage in terms of SNR [38].

In terms of parallel imaging, the most recent work primarily employs either body coils or spine coils as receive coils. The limited number of channels in body coils (only two channels) and the spatial distribution constraints of spine coils result in these setups having either no parallel imaging capability or very limited parallel imaging capability [25, 26, 27, 28]. While previous work has demonstrated that combining NFC coil elements with multi-channel receive coils [32, 39] can achieve parallel imaging, our experimental results in this aspect demonstrate significantly enhanced performance. This suggests that the parallel imaging potential of the NFC coil element can be expanded to multi-element, scalable array applications, which is essential for their implementation in clinical MRI examinations.

It should be noted, however, that the study described in this work has certain limitations. In this study, we used surface coils fixed to the patient table as pickup coils instead of a bore-liner coil array integrated into the magnet to demonstrate the proof of concept. The experiment demonstrates the SNR synthesis principle of the NFC coil system and its potential for parallel imaging. Nevertheless, to fully realize the NFC coil system concept in practical applications, a full-scale experimental model is required. Another limitation of this study is the simple decoupling method used for the NFC coil elements. In NFC coil system design, it is recommended to combine various methods, such as the overlapping approach described in this paper and high-impedance coaxial techniques, to achieve more effective decoupling[40].

## Methods

### SNR analysis

Hoult et al. derived SNR expressions for a pickup coil paired with one or two mutually coupled NFC coils based on Kirchhoff's laws, which are consistent with the traditional expressions [41]. This inspired us to analyze the case of *n* NFC coil elements combined with m pickup coil channels based on Kirchhoff's laws. Differing from their assumption, in this paper, sample noise is considered as a single current loop with resistance $r_o$ and negligible impedance, rather than multiple loops [42].

For an NFC coil system composed of an *m*-channel pickup coil and a *n*-element NFC element, their relationship with sample noise can be expressed in matrix form.

$$\begin{vmatrix} Vin_m \\ \vdots \\ Vin_2 \\ Vin_1 \\ 0 \\ 0 \\ 0 \\ \vdots \\ 0 \end{vmatrix} \cong \begin{vmatrix} Rin_1+r_{m_m}+jX_{m_m} & \cdots & jwM_{m_m m_2} & jwM_{m_m m_1} & jwM_{m_m o} & jwM_{m_m 1} & jwM_{m_m 2} & \cdots & jwM_{m_m n} \\ \vdots & \ddots & & & & & & & \vdots \\ jwM_{m_m m_2} & & Rin_2+r_{m_2}+jX_{m_2} & jwM_{m_1 m_2} & jwM_{m_2 o} & jwM_{m_2 1} & jwM_{m_2 2} & & jwM_{m_2 n} \\ jwM_{m_m m_1} & & jwM_{m_1 m_2} & Rin_3+r_{m_1}+jX_{m_1} & jwM_{m_1 o} & jwM_{m_1 1} & jwM_{m_1 2} & & jwM_{m_1 n} \\ jwM_{m_m o} & & jwM_{m_2 o} & jwM_{m_1 o} & r_o & jwM_{o1} & jwM_{o2} & & jwM_{on} \\ jwM_{m_m 1} & & jwM_{m_2 1} & jwM_{m_1 1} & jwM_{o1} & 1/s_1 & jwM_{12} & & jwM_{1n} \\ jwM_{m_m 2} & & jwM_{m_2 2} & jwM_{m_1 2} & jwM_{o2} & jwM_{12} & 1/s_2 & & jwM_{2n} \\ \vdots & & & & & & & \ddots & \vdots \\ jwM_{m_m n} & \cdots & jwM_{m_2 n} & jwM_{m_1 n} & jwM_{on} & jwM_{1n} & jwM_{2n} & \cdots & 1/s_n \end{vmatrix} \begin{vmatrix} i_{m_m} \\ \vdots \\ i_{m_2} \\ i_{m_1} \\ i_o \\ i_1 \\ i_2 \\ \vdots \\ i_n \end{vmatrix} \equiv V \cong M \cdot I \quad [1]$$

Where $Vin_q$ is the electromotive force (EMF) applied to the q-th pickup coil element, $Rin_q$ is the source impedance and $r_{m_q} + jX_{m_q}$ is the impedance caused by the q-th pickup coil conductor and components on the coil. Clearly, the three subscripts of the mutual inductance coefficient M—$m_q$, o, and p (where p is a natural number between 1 and n)—represent the q-th pickup coil, the equivalent current loop of the sample noise, and the p-th NFC element, respectively. The subscripts of the current $i$ have the same meaning. The subscripted letter $r_o$ represents the resistance of the equivalent current loop of the sample noise, and $s_p$ represents the conductance of the p-th NFC element. The equivalent circuit diagram between a pickup coil element, an NFC coil element, and the sample noise is shown in the Fig. 7.

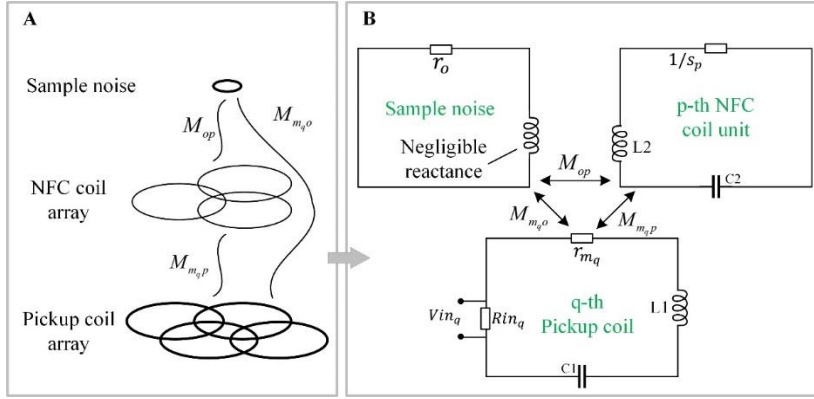

**Figure 7. Pickup coils, NFC coil elements, and sample noise are illustrated.** (A) Mutual inductance coefficient diagram. (B) Equivalent circuit diagram.

To obtain the SNR of the NFC coil system, we need to calculate the total input impedance Zin and the total induced voltage $\xi$ of the pickup coil based on their relationship, since the SNR is defined as [43]:

$$\psi = \frac{\xi}{\sqrt{4kT\Delta\upsilon \operatorname{Re}(Zin)}} \quad [2]$$

In modern coil designs, coupling between channels is typically minimized through the use of overlapping techniques and low-impedance preamplifiers [38]. We assume that each element of the pickup coil is well decoupled. As mentioned earlier, each element of the NFC coil needs to be decoupled to maintain the independence of each element's resonance. Therefore, we consider a well-designed NFC coil to be well-decoupled, too. Accordingly, the M values in the matrix representing the mutual inductance coefficients between pickup elements and between NFC coil elements are all set to zero. The independence between pickup coils allows us to consider separately the interactions between a pickup coil element and the *n* NFC coil elements, before integrating them together.

**Single element pickup coil & NFC coil array**

We consider the pickup coil element indexed as $m_1$ for study, and then the matrix evolved into the format used by Hoult et al.

$$\begin{vmatrix} Vin_1 \\ 0 \\ 0 \\ 0 \\ \vdots \\ 0 \end{vmatrix} \cong \begin{vmatrix} Rin_1+r_{m_1}+jX_{m_1} & jwM_{m_1o} & jwM_{m_11} & jwM_{m_12} & & jwM_{m_1n} \\ jwM_{m_1o} & r_o & jwM_{o1} & jwM_{o2} & & jwM_{on} \\ jwM_{m_11} & jwM_{o1} & 1/s_1 & 0 & & 0 \\ jwM_{m_12} & jwM_{o2} & 0 & 1/s_2 & & 0 \\ & & & & \ddots & \vdots \\ jwM_{m_1n} & jwM_{on} & 0 & 0 & \cdots & 1/s_n \end{vmatrix} \begin{vmatrix} i_{m_1} \\ i_o \\ i_1 \\ i_2 \\ \vdots \\ i_n \end{vmatrix} \equiv V \cong M \cdot I \quad [3]$$

However, it is important to note that they represent different physical meanings, which also leads to different choices in approximation during the derivation of formulas subsequently, resulting naturally in different outcomes. Considering the voltage-current relationship in the loop, the input impedance of the pickup coil is determined to be $Zin_1 = Vin_1/i_{m_1} - Rin_1$. And the above equation can be solved by applying the elimination method to remove currents other than $i_{m_1}$:

$$Zin_1 = \frac{\left[w^2 M_{m_1o}^2 + \sum_{p=1}^n w^2 M_{m_1p}^2 s_p r_o + \sum_{p=1}^n w^4 M_{m_1p}^2 M_{op}^2 s_p - (\sum_{p=1}^n w^2 M_{m_1p} M_{op} s_p)\right] - j[\sum_{p=1}^n 2w^3 M_{m_1o} M_{m_1p} s_p]}{r_o + \sum_{p=1}^n w^2 M_{op}^2 s_p} + (r_{m_1} + jX_{m_1}) \quad [4]$$

Where $r_o$ is the resistance in the sample noise loop that is proportional to the square of the frequency. And terms other than $\sum_{p=1}^n w^2 M_{m_1p}^2 s_p r_o$ can be ignored in the real part of the numerator, since $r_o$ is an enormous value [42]. The minor imaginary part of the numerator can be offset by giving the pickup coil a reactance $jX_{m_1}$ to achieve resonance. In this paper, we ignore the effect of the pickup coil's own resistance losses, as they are relatively small compared to the total input impedance, which is typically 50 ohms [41]. And the input impedance of the q-th pickup coil with subscript $m_1$ under resonance conditions becomes:

$$Zin_1 = \frac{\sum_{p=1}^n w^2 M_{m_1p}^2 s_p r_o}{r_o + \sum_{p=1}^n w^2 M_{op}^2 s_p} \quad [5]$$

To determine the strength of the received signal, an additional ideal current loop parallel to the NFC coil can be introduced, where its cross-sectional area $\delta A$, the current $i_x$ and the magnetic moment $\delta m$ are related by the equation $\delta m = i_x \cdot \delta A$. Let $Rin_{m_1}$ approach infinity, then the new equation can be established as:

$$\begin{vmatrix} 0 \\ Vin_1 \\ 0 \\ 0 \\ 0 \\ \vdots \\ 0 \end{vmatrix} \cong \begin{vmatrix} Rin_1+r_{m_1}+jX_{m_1} & jwM_{m_1x} & jwM_{m_1o} & jwM_{m_11} & jwM_{m_12} & & jwM_{m_1n} \\ jwM_{m_1x} & Rin_x & jwM_{m_1o} & jwM_{m_11} & jwM_{m_12} & & jwM_{m_1n} \\ jwM_{m_1o} & jwM_{xo} & r_o & jwM_{o1} & jwM_{o2} & & jwM_{on} \\ jwM_{m_11} & jwM_{x1} & jwM_{o1} & 1/s_1 & 0 & & 0 \\ jwM_{m_12} & jwM_{x2} & jwM_{o2} & 0 & 1/s_2 & & 0 \\ & & & & & \ddots & \vdots \\ jwM_{m_1n} & jwM_{xn} & jwM_{on} & 0 & 0 & \cdots & 1/s_n \end{vmatrix} \begin{vmatrix} i_{m_1} \\ i_x \\ i_o \\ i_1 \\ i_2 \\ \vdots \\ i_n \end{vmatrix} \equiv V \cong M \cdot I$$

[6]

The EMF induced in the pickup coil can also be determined using the elimination method as:

$$\xi_{m_1} = \frac{[w^2 M_{m_1o} M_{xo} + \sum_{p=1}^{n} w^2 M_{m_1p}^2 M_{xp} s_p r_o + O_R^2] + j[w M_{m_1x} r_o - \sum_{p=1}^{n} w^3 (M_{op}^2 M_{m_1x} - M_{xp} M_{op} M_{m_1o} - M_{m_1p} M_{op} M_{xo}) s_p]}{r_o + \sum_{p=1}^{n} w^2 M_{op}^2 s_p} \cdot i_x \quad [7]$$

Where the term $jwM_{m_1x}r_o$ represents the signal directly induced in the pickup coil, which has a 90° phase difference with the electromotive force induced through the NFC coil. When the distance between the pickup coil and the load exceeds the radius of the pickup coil, it can be ignored [41]. Similar to the input impedance, the voltage $\xi$ induced on the pickup coil can be simplified as:

$$\xi_{m_1} = \frac{\sum_{p=1}^{n} w^2 M_{m_1p}^2 M_{xp} s_p r_o}{r_o + \sum_{p=1}^{n} w^2 M_{op}^2 s_p} \cdot i_x \quad [8]$$

Assuming each NFC coil element has the same conductance $r_w$, the SNR of the NFC coil combined with one pickup coil can be written as:

$$\psi_{m_1} = \frac{wM_0 \delta V \cdot \mathbf{B1}_{m_1}}{\sqrt{n \cdot s_w + \sum_{p=1}^{n} w^2 M_{op}^2 r_o}} \quad [9]$$

Where $n \cdot r_w$ represents the sum of the resistive losses of the n NFC coils, and the term $\sum_{p=1}^{n} w^2 M_{op}^2 r_o$ represents the sample loss proportional to the square of the frequency associated with $n$ NFC elements. $\vec{B1}_{m1}$ is the total field formed by combining the B1 fields of all NFC coil elements when combined with the pickup coil denoted with $m_1$, weighted by $k_{m_p}$.

$$\mathbf{B1}_{m_1} = k_{m_1} \cdot \mathbf{B1}_{w_1} + k_{m_2} \cdot \mathbf{B1}_{w_2} + \ldots + k_{m_n} \cdot \mathbf{B1}_{w_n} \quad [10]$$

Where $\mathbf{B1}_{wp}$ represents the vector field parallel to the XOY plane generated by the NFC coil elements, since $B1_{wp}(r) = M_{xp} / \delta A$, and the ideal current loop denoted by subscript $x$ distributed in space. And $k_{m_p}$ is given by:

$$k_{m_p} = \frac{M_{m_1 p}}{\sqrt{\sum_{p=1}^{n} M_{m_1 p}^2}} \quad [11]$$

Obviously, the coefficient $k_{m_p}$ satisfies $\sum_{p=1}^{n} k_{m_p}^2 = 1$. When there is only one NFC coil, i.e., n=1, the formula reduces to the form presented by Hoult et al [41].

**Pickup coil array & NFC coil array**

At this point, we consider other mutually independent pickup coil channels. At the operating frequency of the 3T magnetic resonance system discussed in this paper, the wavelength of electromagnetic waves in both air and the coaxial line is approximately 2 meters. When combining signals from each pickup coil channel with equal weighting, the phase differences introduced by the varying propagation paths of the channels were ignored in this work. The total EMF signal can thus be expressed as:

$$\xi_{total} = \frac{w^2 r_o (\sum_{p=1}^{n} M_{m_1 p} M_{xp} s_p + \sum_{p=1}^{n} M_{m_2 p} M_{xp} s_p + ... + \sum_{p=1}^{n} M_{m_m p} M_{xp} s_p)}{r_o + \sum_{p=1}^{n} w^2 M_{op}^2 s_p} \qquad [12]$$

While the total input impedance is:

$$Zin_{total} = \frac{w^2 r_o (\sum_{p=1}^{n} M_{m_1 p}^2 s_p + \sum_{p=1}^{n} M_{m_2 p}^2 s_p + ... + \sum_{p=1}^{n} M_{m_m p}^2 s_p)}{r_o + \sum_{p=1}^{n} w^2 M_{op}^2 s_p} \qquad [13]$$

Similar to the SNR collected by a single pickup coil, the total SNR of the NFC coil system according to Equation [9] can be written as:

$$\psi_{total} = \frac{w M_0 \delta V \cdot \mathbf{B1}_{total}}{\sqrt{n \cdot r_w + \sum_{p=1}^{n} w^2 M_{op}^2 s_o}} \qquad [14]$$

Here, $\mathbf{B1}_{total}$ is still the B1 field combined from the $n$ NFC coil elements, however, according to new weight coefficients:

$$\mathbf{B1}_{total} = k_{\eta 1} \cdot \mathbf{B1}_{w1} + k_{\eta 2} \cdot \mathbf{B1}_{w2} + ... + k_{\eta n} \cdot \mathbf{B1}_{wn} \qquad [15]$$

Where

$$k_{\eta p} = \frac{\sum_{q=1}^{m} M_{m_q p}}{\sqrt{\sum_{q=1}^{m} \sum_{p=1}^{n} M_{m_q p}^2}} \qquad [16]$$

It is evident that the coefficient satisfies $\sum_{p=1}^{n} k_{\eta p}^2 = 1$.

The above expression for the signal-to-noise ratio indicates that the distribution of the NFC coil system signal is primarily determined by the weighted combination of the B1 fields of the NFC coils, with the weights depending on the mutual inductance between each NFC coil element and the pickup coil. And the result indicated that when the coefficients of each NFC coil differ significantly, the signal distribution in the same coverage area will be dominated by the B1 field of the NFC coil with the larger coefficient. The conclusion was supported by the previous phantom-based experiments.

Furthermore, when the coefficients for the B1 fields produced by each NFC coil element are identical, the SNR expression becomes essentially the same as that for traditional wired coils. This indicates that the NFC coil system has the potential to surpass the performance of a less optimal rigid coil, as evidenced by the NFC coil system developed in this study.

**Fabrication of NFC coils**

The NFC coils described in this paper were fabricated using flexible printed circuit (FPC) boards with a 50 μm polyimide substrate and 35 μm thick copper. The main trace width is 3 mm, with an overlapping trace width of 0.5 mm between adjacent elements. Each coil element incorporates a variable capacitor (Voltronics, JZ200) with a range of 4.5–20 pF, two diodes (Macom, UM9989), and a 100 μH inductor.

**Bench test**

We used a vector network analyzer (VNA, Keysight E5080A) in conjunction with a pair of mutually decoupled dual-loop detectors to measure the response of the NFC coils to electromagnetic fields (dual-probe method). The VNA's coaxial line ports were configured with the ground ends shorted and the center conductors connected to either end of a detuned circuit to assess its detuning effect (dual-port method). When measuring the coupling or decoupling levels between elements, all coil elements were adjusted to resonance before being disconnected. The two ports of the VNA were connected to the open-circuit ports of the two NFC elements under examination, and the measured S21 characterized the coupling or decoupling level between the two elements. The VNA excitation power was set to -10 dBm.

**3D Electromagnetic Simulations**

We performed 3D electromagnetic simulations using CST, with the simulation dimensions similar to those of the phantom experiment. The conductivity and dielectric constant of the phantom were set to 0.52 and 0.78, respectively.

**MRI sequence parameters for phantom experiments**

All MRI experiments in this study were conducted on a 3T system (uMR 790, Shanghai United Imaging Healthcare, Shanghai, China). In the phantom experiments shown in Figures 2 and 3, we used a phantom (Siemens Inc.) with a diameter of 12 cm and a length of 20 cm, containing a solution with 3.75 g $NiSO_4 \cdot 6H_2O$ and 5 g NaCl per 1000 g of solution. Turbo spin-echo (TSE) sequences were performed with the following parameters: field of view (FOV) = 220 mm × 220 mm, slice thickness = 3 mm, voxel size = 1.72 × 1.72 × 3 mm³, repetition time (TR) = 2439 ms, echo time (TE) = 131.04 ms, bandwidth = 250 Hz/pixel, flip angle (FA) = 90 degrees, and number of averages (NSA) = 1.

For the experiment illustrated in Figure 4, we used a phantom (Shanghai United Imaging Healthcare) with a diameter of 11 cm and a length of 45 cm, containing a solution with 1.243 g $NiSO_4 \cdot 6H_2O$ and 2.6 g NaCl per 1000 g of water. We conducted Turbo spin-echo (TSE) sequences with the following parameters: field of view (FOV) = 220 mm × 220 mm, slice thickness = 3 mm, voxel size = 0.86 × 0.86 × 3 mm³, repetition time (TR) = 5327 ms, echo time (TE) = 78.96 ms, bandwidth = 300 Hz/pixel, flip angle (FA) = 90 degrees, and number of averages (NSA) = 1.

**MRI sequence parameters for in vivo experiments**

In this in vivo study, an adult male volunteer was involved, and the experiment was conducted in accordance with protocols approved by the Human Research Ethics Committee at the Shenzhen Institute of Advanced Technology (SIAT), Chinese Academy of Sciences (CAS), China. In the study, we used spin-echo sequences with the following parameters: field of view (FOV) = 120 mm × 120 mm, slice thickness = 3 mm, voxel size = 0.94 × 0.94 × 3 mm³, repetition time (TR) = 6809 ms, echo time (TE) = 94.2 ms, bandwidth = 250 Hz/pixel, flip angle (FA) = 90 degrees, and number of averages (NSA) = 1 to acquire knee cross-sectional images. The results were used to calculate the SNR and g-factors using the SENSE algorithm. A gradient echo (GRE) sequence with the same parameters was used to acquire noise data. The DICOM images for parallel imaging were obtained using a spin-echo sequence with the following parameters: FOV = 120 mm × 120 mm, slice thickness = 3 mm, voxel size = 1.72 × 1.72 × 3 mm³, TR = 6809 ms, TE = 98.4 ms, bandwidth = 250 Hz/pixel, flip angle (FA) = 90 degrees, and NSA = 1. For high-resolution imaging, the spin-echo sequence was used with the following parameters: FOV = 120 mm × 120 mm, slice thickness = 2 mm, voxel size = 0.33 × 0.33 × 3 mm³, TR = 4436 ms, TE = 139.2 ms, bandwidth = 250 Hz/pixel, auto-calibration signal (ACS) = 2.22, scan time = 2:49 minutes, flip angle (FA) = 90 degrees, and NSA = 1.

## Acknowledgements

This work was supported in part by the Strategic Priority Research Program of Chinese Academy of Sciences, XDB25000000; National Natural Science Foundation of China, U22A20344; Youth Innovation Promotion Association of CAS, Y2021098; Key Laboratory Project of Guangdong Province,

## References:

1. Feinberg DA, *et al.* Next-generation MRI scanner designed for ultra-high-resolution human brain imaging at 7 Tesla. *Nat. Methods* **20**, 2048-2057 (2023).
2. Wu B, *et al.* Flexible transceiver array for ultrahigh field human MR imaging. *Magn. Reson. Med.* **68**, 1332-1338 (2012).
3. Corea JR, *et al.* Screen-printed flexible MRI receive coils. *Nat. Commun.* **7**, 10839 (2016).
4. Li N, *et al.* Simultaneous head and spine MR imaging in children using a dedicated multichannel receiver system at 3T. *IEEE Trans. Biomed. Eng.* **68**, 3659-3670 (2021).
5. Mo Z, *et al.* A novel three-channel endorectal coil for prostate magnetic resonance imaging at 3T. *IEEE Trans. Biomed. Eng.* **70**, 3381-3388 (2023).
6. Li Y, *et al.* Specialized Open-Transmit and flexible receiver Head Coil for High Resolution Ultra-high field fMRI of the Human Somatosensory and Motor Cortex. *Preprint at Research Square* https://www.researchsquare.com /article/rs-4287868/v1 (2024).
7. Port A, *et al.* Detector clothes for MRI: A wearable array receiver based on liquid metal in elastic tubes. *Sci Rep* **10**, 8844 (2020).
8. Zhang B, Sodickson DK, Cloos MA. A high-impedance detector-array glove for magnetic resonance imaging of the hand. *Nat. Biomed. Eng* **2**, 570-577 (2018).
9. Hardy CJ, *et al.* 128-channel body MRI with a flexible high-density receiver-coil array. *J. Magn. Reson. Imaging* **28**, 1219-1225 (2008).
10. Gruber B, Froeling M, Leiner T, Klomp DW. RF coils: A practical guide for nonphysicists. *J. Magn. Reson. Imaging* **48**, 590-604 (2018).
11. Edelstein WA, Glover GH, Hardy CJ, Redington RW. The intrinsic signal-to-noise ratio in NMR imaging. *Magn. Reson. Med.* **3**, 604-618 (1986).
12. Daniel K *et al.* Is a "one size fits all" many-element bore-lining remote body array feasible for routine imaging? *Proc 22th Annual Meeting of ISMRM, Vancouver, Canada*, (2014).
13. Wei J, *et al*. A realization of digital wireless transmission for MRI signals based on 802.11 b. *J. Magn. Reson.* **186**, 358-363 (2007).
14. Aggarwal K, *et al.* A millimeter-wave digital link for wireless MRI. *IEEE Trans. Med. Imaging* **36**, 574-583 (2016).
15. Darnell D, Truong TK, Song AW. Recent advances in radio‐frequency coil technologies: flexible, wireless, and integrated coil arrays. *J. Magn. Reson. Imaging* **55**, 1026-1042 (2022).
16. Nohava L, Ginefri J-C, Willoquet G, Laistler E, Frass-Kriegl R. Perspectives in wireless radio frequency coil development for magnetic resonance imaging. *Front. Physics* **8**, 11 (2020).
17. Wiltshire M, *et al*. Microstructured magnetic materials for RF flux guides in magnetic resonance imaging. *Science* **291**, 849-851 (2001).
18. Luo C, *et al.* Electromagnetic Simulation of Influence of Metamaterial for Magnetic Resonance Imaging at 3T. *Mater Sci Forum* **848,** 347-350 (2016).
19. Zhao X, Duan G, Wu K, Anderson SW, Zhang X. Intelligent metamaterials based on nonlinearity for magnetic resonance imaging. *Adv. Mater.* **31**, 1905461 (2019).

20. Wu K, Zhao X, Bifano TG, Anderson SW, Zhang X. Auxetics‐Inspired Tunable Metamaterials for Magnetic Resonance Imaging. *Adv. Mater.* **34**, 2109032 (2022).
21. Hai A, Spanoudaki VC, Bartelle BB, Jasanoff A. Wireless resonant circuits for the minimally invasive sensing of biophysical processes in magnetic resonance imaging. *Nat. Biomed. Eng* **3**, 69-78 (2019).
22. Alipour A, *et al.* Improvement of magnetic resonance imaging using a wireless radiofrequency resonator array. *Sci. Rep.* **11**, 23034 (2021).
23. Shchelokova A, *et al.* Ceramic resonators for targeted clinical magnetic resonance imaging of the breast. *Nat. Commun.* **11**, 3840 (2020).
24. Zhu X, Wu K, Anderson SW, Zhang X. Helmholtz Coil-Inspired Volumetric Wireless Resonator for Magnetic Resonance Imaging. *Adv. Mater. Technol.* **8**, 2301053 (2023).
25. Chi Z, *et al.* Adaptive cylindrical wireless metasurfaces in clinical magnetic resonance imaging. *Adv. Mater.* **33**, 2102469 (2021).
26. Yi Y, *et al.* In vivo MRI of knee using a metasurface-inspired wireless coil. *Magn. Reson. Med.* **91**, 530-540 (2024).
27. Zhu H, *et al.* Detunable wireless Litzcage coil for human head MRI at 1.5 T. *NMR Biomed.* **37**, e5068 (2024).
28. Wu K, Zhu X, Anderson SW, Zhang X. Wireless, customizable coaxially shielded coils for magnetic resonance imaging. *Sci. Rep.* **10**, eadn5195 (2024).
29. Enhua Xiao *et al.* A novel 8-channel carotid array with two wireless resonator insert for magnetic resonance imaging at 5T. *Proc 32th Annual Meeting of ISMRM, Singapore*, (2024).
30. Zhiguang, et al. Magnetic coupled resonant wireless RF coil for MRI. *Proc 32th Annual Meeting of ISMRM, Singapore*, (2024).
31. Darnell D, Cuthbertson J, Robb F, Song AW, Truong TK. Integrated radio-frequency/wireless coil design for simultaneous MR image acquisition and wireless communication. *Magn. Reson. Med.* **81**, 2176-2183 (2019).
32. Zhu H, *et al.* A detunable wireless resonator insert for high-resolution TMJ MRI at 1.5 T. *J. Magn. Reson.* **360**, 107650 (2024).
33. Lu M, Chai S, Zhu H, Yan X. Low-cost inductively coupled stacked wireless RF coil for MRI at 3 T. *NMR Biomed.* **36**, e4818 (2023).
34. Hoult D, Chen CN, Sank V. Quadrature detection in the laboratory frame. *Magn. Reson. Med.* **1**, 339-353 (1984).
35. Keil B, Wald LL. Massively parallel MRI detector arrays. *J. Magn. Reson.* **229**, 75-89 (2013).
36. Griswold MA, *et al.* Generalized autocalibrating partially parallel acquisitions (GRAPPA). *Magn. Reson. Med.* **47**, 1202-1210 (2002).
37. Hamilton J, Franson D, Seiberlich N. Recent advances in parallel imaging for MRI. *Prog. Nucl. Magn. Reson. Spectrosc.* **101**, 71-95 (2017).
38. Roemer PB, Edelstein WA, Hayes CE, Souza SP, Mueller OM. The NMR phased array. *Magn. Reson. Med.* **16**, 192-225 (1990).
39. Okada T, *et al.* Insertable inductively coupled volumetric coils for MR microscopy in a human 7T MR system. *Magn. Reson. Med.* **87**, 1613-1620 (2022).
40. Li Y, Xie Z, Pang Y, Vigneron D, Zhang X. ICE decoupling technique for RF coil array designs. *Med. Phys.* **38**, 4086-4093 (2011).
41. Hoult D, Tomanek B. Use of mutually inductive coupling in probe design. *Concepts Magn. Resonance* **15**, 262-285 (2002).
42. Edelstein W, Bottomley PA, Pfeifer LM. A signal-to-noise calibration procedure for NMR imaging systems. *Med. Phys.* **11**, 180-185 (1984).

43. Hoult D, Lauterbur PC. The sensitivity of the zeugmatographic experiment involving human samples. *J. Magn. Reson.* **34**, 425-433 (1979).